\documentclass[preprint2]{aastex}

\begin{document}

\title{Long-term optical and X-ray observations of the old novae 
DI~Lacertae and V841~Ophiuchi\altaffilmark{1}}

%\author{D. W. Hoard}
%\affil{Cerro Tololo Inter-American Observatory, Casilla 603, La Serena, Chile}
%\email{dhoard@noao.edu}
%\authoraddr{CTIO, P.O. Box 26732, Tucson AZ 85726}
%
%\author{Paula Szkody}
%\affil{Department of Astronomy, University of Washington, Box 351580, 
%Seattle WA 98195-1580}
%\email{szkody@astro.washington.edu}
%
%\author{R. K. Honeycutt}
%\affil{Astronomy Department, Indiana University, Swain Hall West 319, %Bloomington IN 47505}
%\email{honey@astro.indiana.edu}
%
%\author{Jeff Robertson}
%\affil{Department of Physical Sciences, Arkansas Technical University, Russellville AR 72801-2222}
%\email{Jeff.Robertson@mail.atu.edu}
%
%\author{Vandana Desai}
%\affil{Department of Astronomy, University of Washington, Box 351580, 
%Seattle WA 98195-1580}
%\email{desai@astro.washington.edu}
%\and
%\author{T. Hillwig}
%\affil{Astronomy Department, Indiana University, Swain Hall West 319, 
%Bloomington IN 47505}
%\email{thillwig@astro.indiana.edu}

\author{D. W. Hoard\altaffilmark{2}, Paula Szkody\altaffilmark{3}, 
R. K. Honeycutt\altaffilmark{4}, Jeff Robertson\altaffilmark{5}, 
Vandana Desai\altaffilmark{3}, \\ and~T. Hillwig\altaffilmark{4}}

\altaffiltext{1}{Based on observations with the Apache Point Observatory 
3.5-m telescope which is owned and operated by the Astrophysical 
Research Consortium, and on observations with the WIYN Observatory 
3.5-m telescope which is jointly operated by the University of Wisconsin, 
Indiana University, Yale University, and the National Optical Astronomy 
Observatories.}
\altaffiltext{2}{Cerro Tololo Inter-American Observatory, Casilla 603, 
La Serena, Chile, {\tt dhoard@noao.edu}}
\altaffiltext{3}{Department of Astronomy, University of Washington, 
Box 351580, Seattle, WA 98195-1580}
\altaffiltext{4}{Astronomy Department, Indiana University, Swain Hall 
West 319, Bloomington, IN 47505}
\altaffiltext{5}{Department of Physical Sciences, Arkansas Technical 
University, Russellville, AR 72801-2222}

\begin{abstract}
We present an analysis of ground-based optical photometry and 
spectroscopy, and {\em Rossi X-ray Timing Explorer} X-ray observations 
of the old novae DI Lacertae and V841 Ophiuchi.  Our optical 
photometry data (obtained with the automated photometry telescope 
RoboScope) comprise an almost decade-long light curve for each 
star, while the contemporaneous spectroscopy and X-ray observations 
repeatedly sampled each nova during separate intervals of 
$\approx45$--55 d in length.  The long-term optical light curves 
of both novae reveal quasiperiodic variability on typical time scales of 
$\sim30$--50 d with amplitudes of $\Delta V\sim0.4$--0.8 mag.  
V841 Oph also displays a long-term, sinusoidal modulation of its 
optical light on a time scale of 3.5--5 yr.  The optical spectra 
of these novae display quite different characteristics from each 
other, with DI Lac showing narrow Balmer emission cores situated 
in broad absorption troughs while V841 Oph exhibits strong 
single-peaked Balmer, \ion{He}{1} and \ion{He}{2} emission lines.  
We find little change between spectra obtained during different 
optical brightness states.  The X-ray count rates for both novae 
were very low ($\lesssim1.5$ ct s$^{-1}$) and there was no reliable 
correlation between X-ray and optical brightness.  The combined 
X-ray spectrum of DI Lac is best fit by a bremsstrahlung emission 
model (with $kT\sim4$ keV and $N_{\rm H} < 1.8\times10^{22}$ cm$^{-3}$); 
the X-ray spectrum of V841 Oph is too weak to allow model fitting.  
We discuss the possible origin of variability in these old novae in 
terms of magnetic activity on the secondary star, dwarf nova type disk instabilities, and the ``hibernation'' scenario for cataclysmic 
variable stars. 

{\tt Accepted by PASP on 28 August 2000 for the December 2000 issue.} 
\end{abstract}

%These are ApJ, MNRAS, A&A keywords

\section{Introduction}
\label{s-intro}

Cataclysmic variables (CVs) are semi-detached interacting binary 
stars composed of a white dwarf (WD) primary star and a low mass 
($\lesssim0.5$M$_{\odot}$) main sequence secondary star, with 
typical orbital periods of $\lesssim1$ d.  The Roche-lobe-filling 
secondary star loses mass through the inner Lagrangian point into 
an accretion disk formed around the WD (for non-magnetic CVs).  
Classical novae are a subclass of CV in which a thermonuclear 
runaway is triggered in a reservoir of matter that has been gradually 
accreted onto the WD.  The resultant outburst produces a peak 
brightness increase of $\approx$ 6--15 mag, and releases 
$10^{44}-10^{46}$ erg \cite[see review in][ch.\ 5]{Warn95}. 

The outbursts of novae are often well-observed, and the long term 
behavior of many dwarf novae and novalike CVs is monitored by the 
American Association of Variable Star Observers.  However, there 
have been few programs for monitoring the variability of novae at 
quiescence.  One such program, using an automated telescope called 
RoboScope (see \S\ref{s-robo}), has monitored 22 old novae and 42 
novalikes, most of them for over 9 years now.  RoboScope has found 
several kinds of unusual photometric behavior.  About one-third of 
the old novae have shown quasi-periodic variability for a year or 
two, interspersed with stable light curves. These variations do not 
appear stochastic as they repeat at similar periods when they reappear. 
Preliminary analysis of RoboScope light curves spanning several years 
for the old nova DI Lacertae suggested brightness oscillations on a 
time scale of $\approx35$ d \citep{Hone95}, while the old nova 
V841 Ophiuchi also displayed prominent variability \citep{Hone94}.
In order to determine if these oscillations could be related to a 
disk instability mechanism (as operates in dwarf novae) or to mass 
transfer or magnetic effects on the secondary star, we attempted a 
detailed study of DI Lac and V841 Oph, which are among the most 
active and brightest of the systems showing the oscillations.

DI Lac (= Nova Lac 1910) was a moderately fast nova that reached a 
maximum brightness of $m_{V}({\rm max}) = 4.3$ mag during its outburst.  
V841 Oph (= Nova Oph 1848) was a slow nova that reached a similar 
brightness, $m_{V}({\rm max}) = 4.2$ mag \citep{Kuka71}.
The reddening for both objects has been measured from 
{\em International Ultraviolet Explorer} ($IUE$) ultraviolet and 
ground-based optical spectra.  The range for DI Lac 
is $E(B-V)=0.15$--0.41 \citep{Bruc84,Cass90}, while for V841 Oph 
it is $E(B-V)=0.30$--0.58 \citep{Bruc84,Cass90,Weig94,Verb97}.
Nova shells were not detected in H$\alpha$ for either 
CV \citep{Cohe85}.  In X-rays, DI Lac was undetected in the $ROSAT$ 
PSPC All Sky Survey, with a $2\sigma$ upper limit of 0.013 cts s$^{-1}$ 
in the 11--201 channel (0.1--2.0 keV) energy range during 299 s total 
exposure time, while V841 Oph was a marginal detection with 
$0.017\pm0.007$ cts s$^{-1}$ in the ``hard'' 0.9--2.0 keV {\em ROSAT} 
energy range ($0\pm0.025$ cts s$^{-1}$ in the full {\em ROSAT} energy 
range) during 402 s total exposure time \citep{Verb97}.

We report here on the results of ground-based optical photometry and 
spectroscopy, and {\em Rossi X-ray Timing Explorer} ($RXTE$) X-ray 
observations of DI Lac and V841 Oph.  The optical photometry comprises 
an almost decade-long light curve for each star, while the 
contemporaneous spectroscopy and X-ray observation repeatedly sampled 
each nova during separate intervals of $\approx45$--55 d in length.

\section{Observations and Analysis}

Our long-term optical photometry of DI Lac and V841 Oph was accomplished 
from 1990--1998, while the spectroscopic amd X-ray observations took 
place in 1997.  These observations are summarized in Tables~\ref{t-DI_log} 
and \ref{t-V8_log}, and are discussed in detail below.

\subsection{Optical Photometry}
\label{s-robo}

\begin{figure}[tb]
\epsscale{1.00}
\plotone{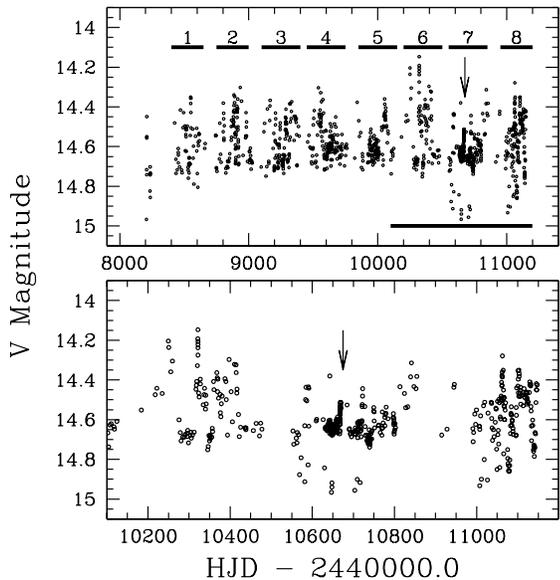}
\epsscale{1.00}
\caption{The optical ($V$) light curve of DI Lac from RoboScope.  The 
top panel shows the entire light curve, while the bottom panel shows 
the region around the $RXTE$ observations. The mid-point time of the 
$RXTE$ observations is marked with an arrow.  The horizontal bars 
above the data in the top panel show the eight observing seasons (the 
short initial observing season around HJD 2448200 is not marked), 
while the bar below the data shows the range of HJD displayed in the 
bottom panel.\label{f-DI_lc}}
\end{figure}

\begin{figure}[tb]
\epsscale{1.00}
\plotone{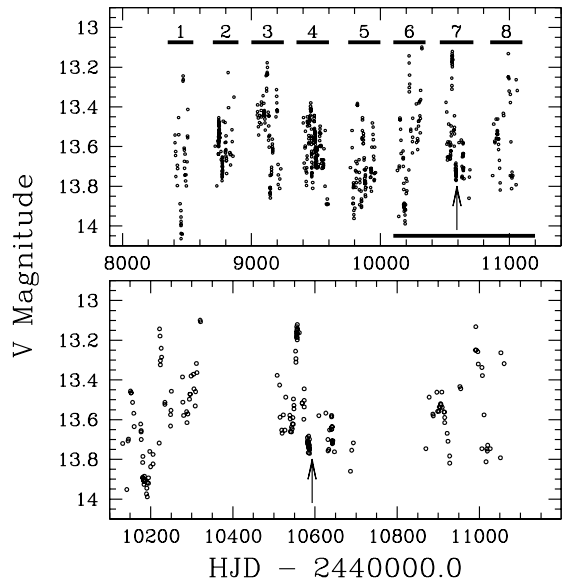}
\epsscale{1.00}
\caption{As in Figure~\ref{f-DI_lc}, but for V841 Oph.\label{f-V8_lc}}
\end{figure}

Our optical photometry data were acquired by RoboScope, 
a 41-cm telescope in Indiana equipped for automated differential 
CCD stellar photometry \citep{HT92}.  All observatory operations 
(including data reductions) are accomplished
as fully unattended and unsupervised tasks,
which makes practical the acquisition of long homogeneous data streams. 
Typically, RoboScope obtains one or two 4-min exposures per clear
night for each of $\sim140$ program stars.  The data are reduced using the
method of incomplete ensemble photometry \citep{Hone92}.  For DI Lac, 
85 ensemble stars were used in 987 exposures over 9 observing seasons
from 1990 November to 1998 November.  
For V841 Oph, 24 ensemble stars were used in
611 exposures over 8 observing seasons from 1991 May to 1998 September.
The zero-points of the differential light curves were established using 
secondary standard stars from \citet{Hend95,Hend97}.  Six such
secondary standards were used for V841 Oph, while 13 were employed for DI Lac.  
Typical $1\sigma$ uncertainties in the calibrated magnitudes are on the 
order of 0.01--0.05 mag.  Over the entire multiyear range of the RoboScope 
light curves, the mean magnitudes were $V\approx14.55$ for DI Lac 
and $V\approx13.6$ for V841 Oph, with full ranges of variability of 
$\Delta V\approx0.9$ mag and $\Delta V\approx1$ mag, respectively.  
The RoboScope light curves of DI Lac and V841 Oph are shown in 
Figures \ref{f-DI_lc} and \ref{f-V8_lc}.

%%%%%%%%%%%%%%%%%%%%%%%%%%%%%%%%%%%%%%%%%%%%%%%%%%%%%%%%%%%%%%%%%
% cp /uw62/hoard/Stars/DI_Lac/lc/lc_both.eps fg1.eps
%%BoundingBox: 30 160 530 690
%%%%%%%%%%%%%%%%%%%%%%%%%%%%%%%%%%%%%%%%%%%%%%%%%%%%%%%%%%%%%%%%%%%
%%%%%%%%%%%%%%%%%%%%%%%%%%%%%%%%%%%%%%%%%%%%%%%%%%%%%%%%%%%%%%%%%
% cp /uw62/hoard/Stars/V841_Oph/lc/lc_both.eps fg2.eps
%%BoundingBox: 30 160 530 690
%%%%%%%%%%%%%%%%%%%%%%%%%%%%%%%%%%%%%%%%%%%%%%%%%%%%%%%%%%%%%%%%%%%

With only limited excursions to anomalously fainter magnitudes, the 
typical minimum magnitude of $\approx14.7$ for DI Lac has a sharp 
boundary at any given epoch.  However, the minimum brightness displays 
a slight, apparently linear, trend towards increasing brightness over 
the 8 years of RoboScope observation.  If we ignore for the moment the 
small number of data points that fall below the well-defined low 
brightness limit seen in the top panel of Figure \ref{f-DI_lc}, then 
the typical minimum brightness of DI Lac changes from $V\approx14.75$ mag 
at the beginning of the RoboScope coverage, to $V\approx14.67$ mag at 
the end.  This corresponds to a mean increase of the minimum brightness 
by $\Delta V\approx -0.01$ mag yr$^{-1}$.  The maximum brightness of 
DI Lac is much less uniform, and it is difficult to determine if the 
maximum brightness behaves in the same manner as the minimum 
brightness.  A linear fit to the complete DI Lac light curve (with 
each data point weighted by the inverse square of its $1\sigma$ 
uncertainty) gives the rate of change of the mean magnitude 
as $< +0.001$ mag yr$^{-1}$; that is, the mean brightness of DI Lac 
is essentially constant.  However, if we exclude the 2 excursions 
to fainter magnitudes at HJD 2450600 and HJD 2451000 (as well as 
the excursion to brighter magnitudes at HJD 2450300), then the slope 
of the linear fit changes to $\Delta V = -0.004$ mag yr$^{-1}$; that 
is, a trend towards increasing mean brightness.  

The faint magnitude excursions in the light curve of DI Lac are 
very similar to the ``dips'' seen by \citet{Hone98} in long-term 
RoboScope light curves of five old novae and novalike CVs.  In the 
latter systems, the dips were found to often be paired with a 
preceding or following outburst, although the dips also sometimes 
occurred as isolated events.  \citet{Hone98} do not find a clear 
mechanism (e.g.\ disk instabilities, truncated disks, mass transfer 
modulation) that is responsible for the dips, but conclude that the 
overall photometric behavior (dips and outbursts) of the novae and 
novalikes that they studied is likely to be governed by either a 
combination of disk and mass transfer events, or by mass transfer 
events alone.

In contrast to the linear trend in the minimum brightness of DI Lac, 
the minimum brightness of V841 Oph appears to follow an almost 
sinusoidal trend that completes approximately 1.5--2 cycles during 
the 7.5 year RoboScope coverage.  As seen in the top panel of 
Figure \ref{f-V8_lc}, the minimum brightness of V841 Oph varies 
from $V\approx14.5$ mag to $V\approx13.8$ mag.  The maximum 
brightness of V841 Oph, which ranges from $V\approx13.1$ mag to 
$V\approx13.4$ mag, displays the same behavior; that is, when 
the minimum brightness at a given epoch is at its faint (bright) 
level, the maximum brightness is also at its faint (bright) level.  
We note that the magnitude range spanned by the minimum brightness 
in V841 Oph is about twice as large as that spanned by the maximum 
brightness.  Although a linear fit to this overall trend is perhaps 
not the optimum choice, the corresponding rate of mean magnitude 
change is $\Delta V = +0.010$ mag yr$^{-1}$.  It is possible (but 
not required by the data!) that the apparently linear trend in the 
light curve of DI Lac may also be a sinusoid, but with a much longer 
cycle length than in V841 Oph.

\begin{figure}[tb]
\epsscale{1.00}
\plotone{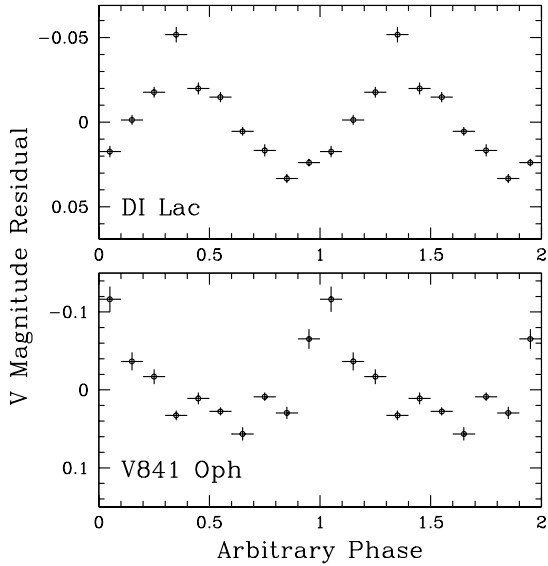}
\epsscale{1.00}
\caption{The RoboScope light curves of DI Lac (top panel) and 
V841 Oph (bottom panel) folded on periods of 37 d and 36 d, 
respectively.  A best-fit linear trend was subtracted prior to 
folding.  The data have been averaged into phase bins of width 
0.1, and are repeated twice.  The horizontal error bars show the 
width of the phase bins, while the vertical error bars span 
$\pm3\sigma_{\rm n}$, where $\sigma_{\rm n}$ is the standard 
deviation of the mean in each phase bin.\label{f-folded_lc}}
\end{figure}

We searched for periodicities in the range 10--100 days in the 
RoboScope light curve data in two ways: by applying the phase 
dispersion minimization (PDM) algorithm \citep{Stel78}, as well 
as performing an independent power spectrum analysis using the 
{\sc clean} algorithm \citep{Robe87}.  We performed the period 
search on both the entire data set for each CV, as well as on 
the data subsets from individual observing seasons (indicated 
in Figures \ref{f-DI_lc} and \ref{f-V8_lc}).  The most prominent 
period detected in each data set is listed in Table \ref{t-periods} 
(in several cases, most notably for the combined data sets for 
each nova, two significant periods are listed).  
In general, the two period search methods gave equivalent results 
to within better than 1--4 days.
All of the detected time scales appear to correspond to 
quasiperiodic behavior rather than truly periodic variability 
such as that resulting from, for example, an eclipsing orbit.  
The time scales have significant ``flexibility'' (on the order 
of 1--5 days for periods $\lesssim50$ d, 
5--20 days for periods $\gtrsim50$ d) over the multi-year 
length of the light curves (as 
shown by substantially broadened dips and peaks in the PDM and 
{\sc clean} analyses, respectively).  Phase-binned light curves 
of both CVs folded on the shorter period found for each combined data 
set are shown in Figure \ref{f-folded_lc}.  
We note that the true amplitude of variation in a given cycle or 
observing season will be larger than the 0.1--0.2 mag suggested 
by Figure \ref{f-folded_lc}, in which data from many cycles with 
slightly different amplitudes and periods have been averaged 
together.  Inspection of Figures \ref{f-DI_lc} and \ref{f-V8_lc} 
shows amplitudes of 0.4--0.8 mag for the full range of variability 
within each observing season.  The long-term sinusoidal trend in 
the light curve of V841 Oph is also well-fit by a periodicity in 
the range 1800--1900 d or, with slightly less agreement, 1250--1300 d.  
(We note that these two period ranges are approximately related by 
the ratio 3:2, so we are likely seeing an aliasing effect due to the 
fact that only 1.5--2 cycles of these long periods are contained in 
the light curves.)

%%%%%%%%%%%%%%%%%%%%%%%%%%%%%%%%%%%%%%%%%%%%%%%%%%%%%%%%%%%%%%%%%
% cp /uw62/hoard/Stars/DI_Lac/lc/clean/fold.eps fg3.eps
%%BoundingBox: 30 165 540 685
%%%%%%%%%%%%%%%%%%%%%%%%%%%%%%%%%%%%%%%%%%%%%%%%%%%%%%%%%%%%%%%%%%%

\subsection{Optical Spectroscopy}
\label{s-optspec}

We obtained optical spectra of both CVs using the Double Imaging 
Spectrograph on the Apache Point Observatory (APO) 3.5-m 
telescope \citep{Gill95}\footnote{also see \url{http://www.apo.nmsu.edu/}} 
during University of Washington share time.  Spectra of DI Lac were 
obtained in July and August of 1997; spectra of V841 Oph were 
obtained in May of 1997.  The APO spectra have a resolution of 
$\approx2$\AA\ and cover simultaneous wavelength ranges 
of $\approx4200$--5000\AA\ and $\approx5800$--6800\AA.  The raw 
spectrum images were reduced in the standard fashion using IRAF.  
The instrumental response was removed via spectra of standard 
stars \citep{Mass88} obtained on the same nights; however, slit 
losses due to guiding errors rendered the absolute flux calibration 
unreliable.  Consequently, we have normalized the spectra to a 
constant continuum level of 1.0.    

\begin{figure}[tb]
\epsscale{1.00}
\plotone{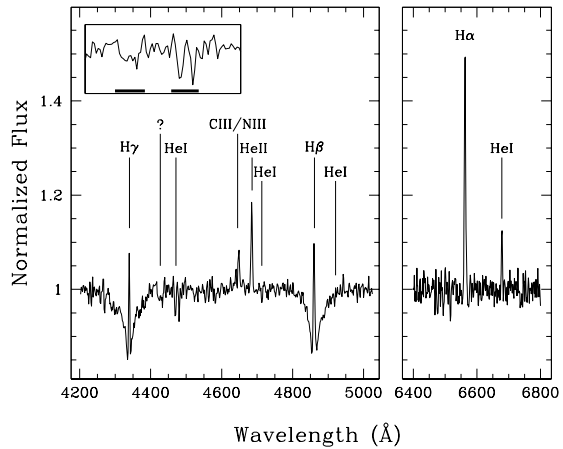}
\epsscale{1.00}
\caption{Representative optical spectrum of DI Lac from APO on 
1997 Jul 25 09:01:47 UT (HJD 2450654.87744).  The spectrum has 
been normalized to a continuum level of 1.0.  The wavelengths 
of prominent line transitions are marked.  The box inset in 
the left panel shows an expanded view of the region from 4390\AA\ 
to 4520\AA.  The two features highlighted by dark underlines are 
marked (from left to right) ``?'' and ``HeI'' in the full 
spectrum plot.\label{f-DI_spec}}
\end{figure}

A representative spectrum of DI Lac is shown in Figure \ref{f-DI_spec}. 
%%%%%%%%%%%%%%%%%%%%%%%%%%%%%%%%%%%%%%%%%%%%%%%%%%%%%%%%%%%%%%%%%
% cp /uw62/hoard/Stars/DI_Lac/APO/spec.eps fg4.eps
%%BoundingBox: 10 175 590 660
%%%%%%%%%%%%%%%%%%%%%%%%%%%%%%%%%%%%%%%%%%%%%%%%%%%%%%%%%%%%%%%%%%%
Note the prominent Balmer absorption troughs containing narrow, 
central emission cores.   Our other two APO spectra of DI Lac 
(not shown) are essentially identical to this one, with the 
exception that the Balmer emission cores in the August spectrum 
are somewhat stronger (relative to \ion{He}{2}) than in the July 
spectra.  The \ion{He}{1} $\lambda4471$ feature (see inset) is 
present with the same profile shape in all three of our spectra.  
It is suggestive of the absorption trough with central emission 
as seen in the Balmer lines.   The feature labeled with ``?'' is 
likely an instrumental or reduction artifact since its wavelength 
does not match any identifiable line and it occurs in {\em all} of 
our APO spectra (those of V841 Oph included).

\begin{figure}[tb]
\epsscale{1.00}
\plotone{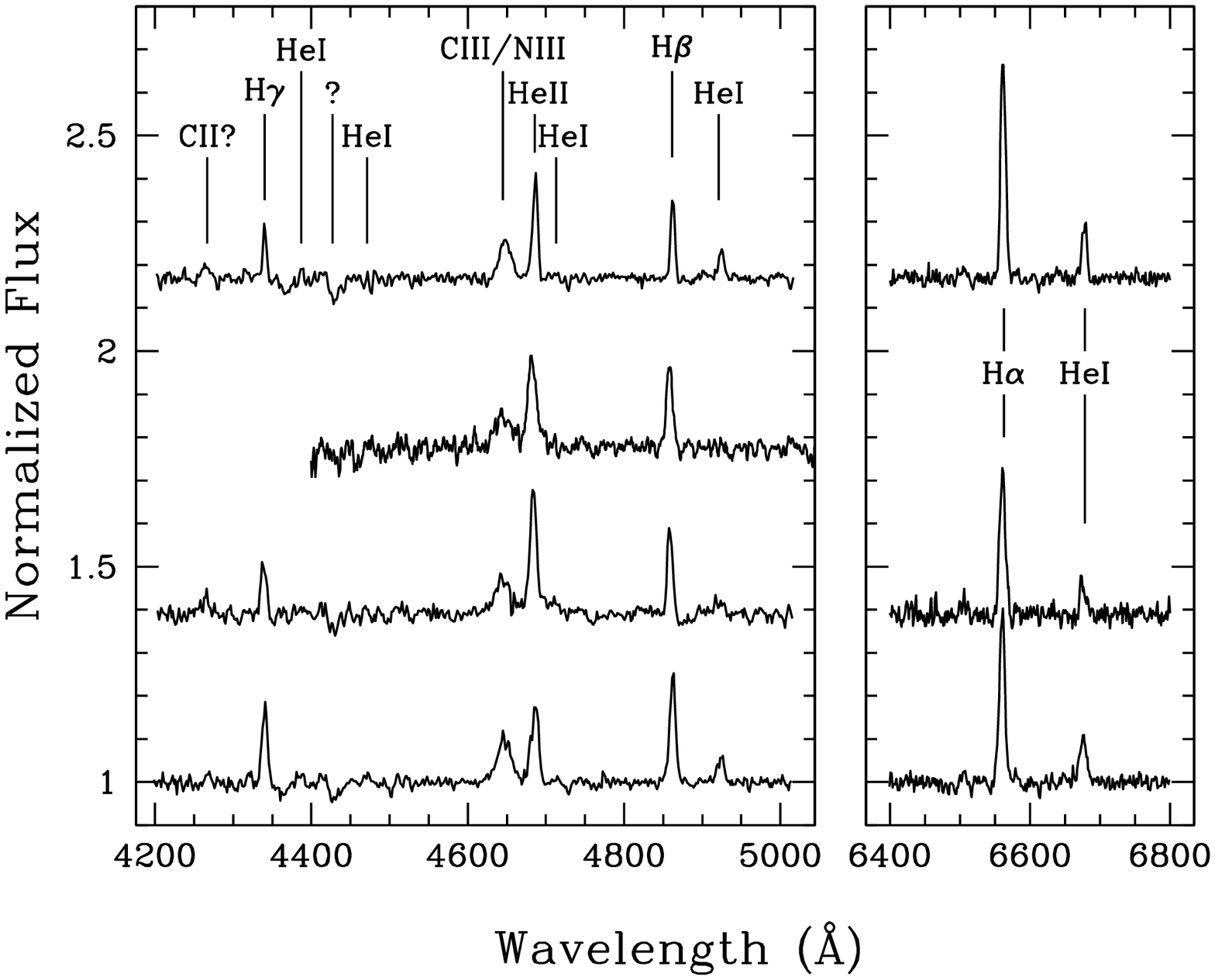}
\epsscale{1.00}
\caption{Optical spectra of V841 Oph on (bottom to top): 1997 May 07 UT 
(APO), 1997 May 15 UT (APO), 1997 May 17 UT (WIYN), and 1997 May 27 UT 
(APO).  The spectra have been normalized to a continuum level 
of 1.0.\label{f-V8_spec}}
\end{figure}

Additional optical spectra of both novae were obtained using the 
Hydra Multi-fiber Positioner + Bench Spectrograph \citep{Bard92,Bard94} 
on the WIYN 3.5-m 
telescope\footnote{see \url{http://www.noao.edu/wiyn/}}.  
Spectra of DI Lac were obtained in 1997 September, and spectra of 
V841 Oph were obtained in 1997 May.  The WIYN spectra have a 
resolution of $\approx1$\AA\ and cover (usable) wavelength ranges 
of $\approx6200$--6800\AA\ for DI Lac and $\approx4400$--5000\AA\ 
for V841 Oph.  For the WIYN spectra of DI Lac, no standard star 
observations were available, so it was not possible to correct for 
the instrumental response or perform a flux calibration.  We do 
not show these spectra here, but note that they are qualitatively 
similar to the red APO spectrum shown in Figure \ref{f-DI_spec} -- no 
differences corresponding to the different brightness levels in the 
RoboScope light curve are apparent.  The V841 Oph WIYN data were 
reduced and calibrated as with the APO spectra.  Again, slit losses 
and, additionally, the presence of cirrus made the absolute flux 
calibration unreliable, so we have normalized the spectra to a 
constant continuum level of 1.0.  All of the spectra of V841 Oph 
are shown in Figure \ref{f-V8_spec} (the three WIYN spectra have 
been averaged together to improve the S/N).  
%%%%%%%%%%%%%%%%%%%%%%%%%%%%%%%%%%%%%%%%%%%%%%%%%%%%%%%%%%%%%%%%%
% cp /uw62/hoard/Stars/V841_Oph/spectra/spec.eps fg5.eps
%%BoundingBox: 10 175 590 660
%%%%%%%%%%%%%%%%%%%%%%%%%%%%%%%%%%%%%%%%%%%%%%%%%%%%%%%%%%%%%%%%%%%
There is no indication of the absorption troughs seen in the spectra 
of DI Lac.  Additionally, the \ion{C}{3}/\ion{N}{3} + \ion{He}{2} 
emission complex appears to be stronger, with broader lines, in V841 Oph 
than in DI Lac.

Flux-calibrated optical spectra of both DI Lac and V841 Oph are 
shown in \citet{Will83}, with resolutions of 5.1\AA\ and 6.7\AA, 
respectively, and in \citet{Ring96}, with resolution of 15--20\AA\ for 
both.  \citet{Will83} obtained his spectra in 1981, while \citet{Ring96} 
obtained theirs in 1990.  In both papers, the spectra of both novae 
show steep blue continua; the observed continuum of DI Lac is bluer 
than that of V841 Oph (but different reddening values could 
eliminate or reverse this characteristic).  \citet{Ring96} fit 
a power law of the form 
$F_{\lambda} \propto \lambda^{-\alpha}$ 
to the continuum regions 5600--6500\AA\ and 6650--8000\AA\ in their 
spectra.  These fits yielded indices of $\alpha=1.81(2)$ -- dereddened 
to $\alpha_{0}=3.10(3)$ using $E(B-V)=0.41$ -- for DI Lac 
and $\alpha=1.45(1)$ -- dereddened to $\alpha_{0}=2.79(1)$ 
using $E(B-V)=0.58$ -- for V841 Oph.   
\citet{Cass89} performed this power law fit to the wavelength 
region 1200--3200\AA\ in $IUE$ UV spectra of DI Lac and V841 Oph.  
Using reddenings at the low ends of the ranges given 
in \S\ref{s-intro}, $E(B-V)=0.15$ for DI Lac and $E(B-V)=0.30$ 
for V841 Oph, they calculated dereddened slopes of $\alpha_{0}=1.5$ 
and $\alpha_{0}=2.0$, respectively.  Comparison with the slope 
expected for the continuum flux distribution of a steady-state 
accretion disk composed of parcels radiating as 
blackbodies, $\alpha=2.3$ \citep{Prin81}, suggests that the true 
reddening values for DI Lac and V841 Oph are near the middle of 
the ranges given in \S\ref{s-intro}.

In 1936, both of these novae were observed spectroscopically 
by \citet{Huma38}, who noted their strong blue continua, but 
also noted the absence of {\em any} lines, emission or absorption.  
Since that time, the available published spectra of DI Lac and 
V841 Oph in \citet{Gree60}, \citet{Will83}, and \citet{Ring96} 
are consistent with our current spectra \cite[see spectroscopic 
histories summarized in][]{Ring96}, with the exception 
that \citet{Will83} did not report the detection of 
any \ion{He}{2} $\lambda4686$ emission in either star.   

\subsection{X-ray Observations}

DI Lac and V841 Oph were each observed ten times during 
July--September and April--June of 1997, respectively, with 
the Proportional Counter Array \cite[PCA;][]{Jaho96} on 
the $RXTE$ satellite \cite[e.g.][]{Brad93}\footnote{also 
see \url{http://heasarc.gsfc.nasa.gov/docs/xte/}}. 
The PCA consists of five xenon-methane proportional counters 
effective over the range 2--60 keV (with 18\% energy resolution 
at 6 keV).  Each $RXTE$ visit constituted a Good-Time-Interval 
on-source lasting $\approx1000$--2500 s.  We extracted the net 
average X-ray count rate from each visit and the X-ray spectrum 
from the ten combined visits to each star using 
the FTOOLS/LHEASOFT (v5.0) software package.  We utilized the 
faint source background models (pca\_bkgd\_faintl7\_e3v19990824.mdl 
and pca\_bkgd\_faint240\_e3v19990909.mdl) and an up-to-date 
South Atlantic Anomaly history file (pca\_saa\_history\_20000223).  
Extraction of the count rates and spectra was carried out as 
described in ``The RXTE Cookbook''\footnote{
\url{http://heasarc.gsfc.nasa.gov/docs/xte/recipes/cook\_book.html}}.

\begin{figure}[tb]
\epsscale{1.00}
\plotone{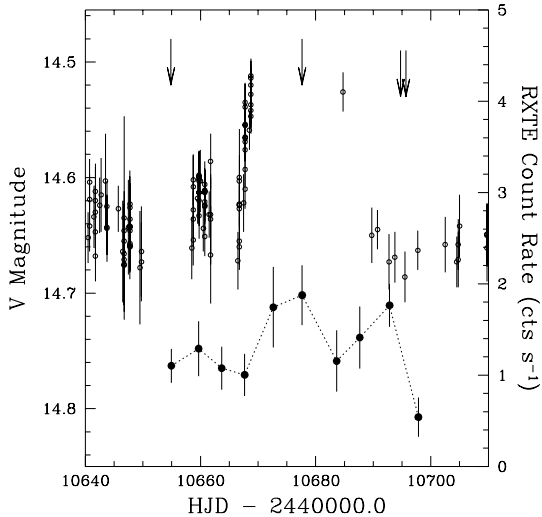}
\epsscale{1.00}
\caption{The optical ($V$) light curve of DI Lac from RoboScope 
(small circles, left axis scale) with the mean $RXTE$ count rate 
(0--15 keV) at each of the ten epochs of observation (large circles 
connected by a dotted line, right axis scale).  One-sigma error 
bars are shown on all of the points; the arrows mark the times 
when optical spectra were obtained.  The lowered arrows mark the 
WIYN spectra; the others are from APO.\label{f-DI_Xray}}
\end{figure}

\begin{figure}[tb]
\epsscale{1.00}
\plotone{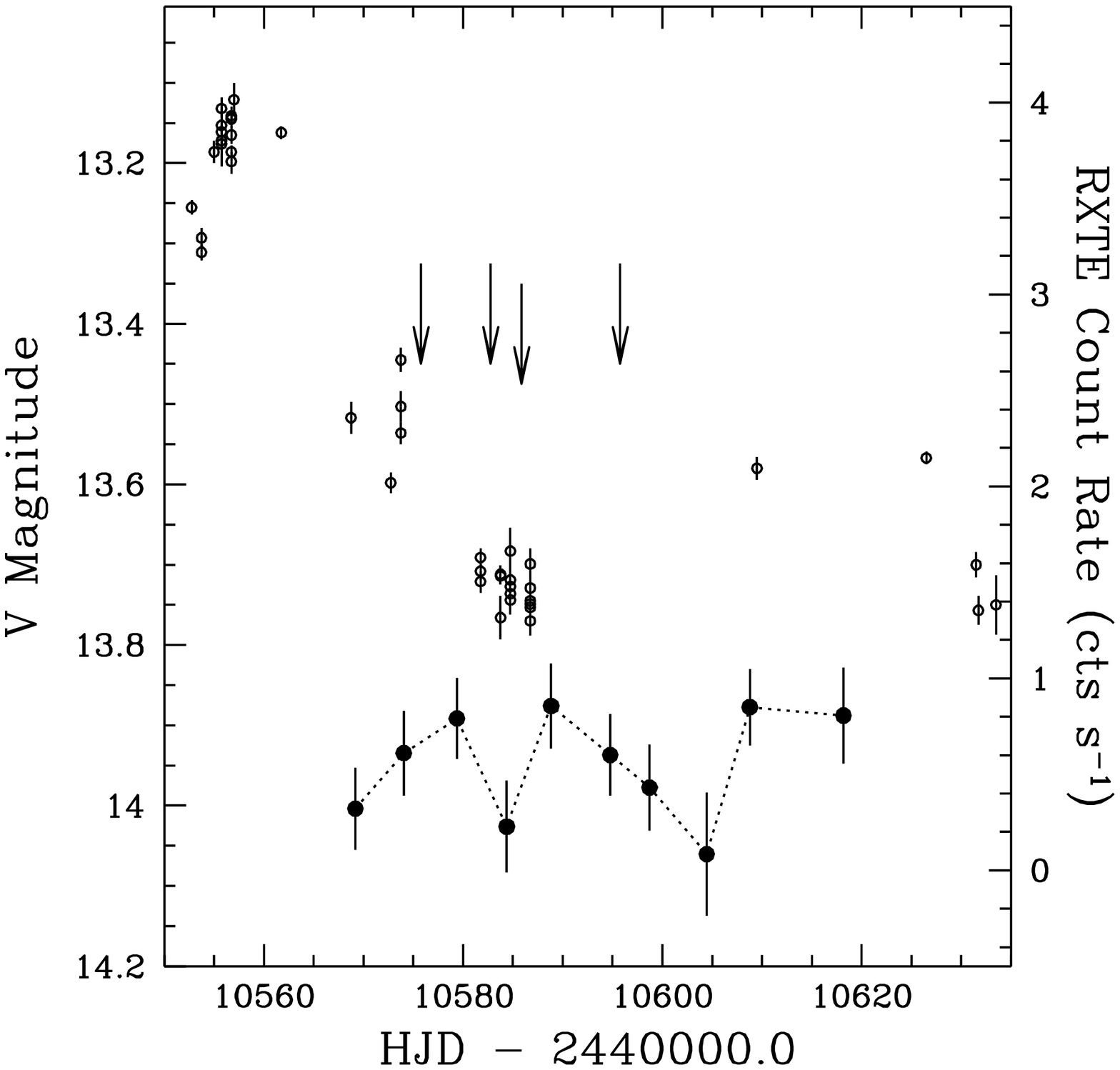}
\epsscale{1.00}
\caption{As in Figure~\ref{f-DI_Xray}, but for V841 Oph.\label{f-V8_Xray}}
\end{figure}

The X-ray light curves constructed from the mean count rates at 
each visit for DI Lac and V841 Oph are shown in 
Figures \ref{f-DI_Xray} and \ref{f-V8_Xray}, respectively, 
along with the corresponding sections of their RoboScope $V$ band 
light curves.
%%%%%%%%%%%%%%%%%%%%%%%%%%%%%%%%%%%%%%%%%%%%%%%%%%%%%%%%%%%%%%%%%
% cp /uw62/hoard/Stars/DI_Lac/lc/lc.eps fg6.eps
%%BoundingBox: 30 155 590 695
%%%%%%%%%%%%%%%%%%%%%%%%%%%%%%%%%%%%%%%%%%%%%%%%%%%%%%%%%%%%%%%%%%%
%%%%%%%%%%%%%%%%%%%%%%%%%%%%%%%%%%%%%%%%%%%%%%%%%%%%%%%%%%%%%%%%%
% cp /uw62/hoard/Stars/V841_Oph/lc/lc.eps fg7.eps
%%BoundingBox: 30 155 590 695
%%%%%%%%%%%%%%%%%%%%%%%%%%%%%%%%%%%%%%%%%%%%%%%%%%%%%%%%%%%%%%%%%%%
The light curve counts were summed in the energy range 0--15 keV.  
Inspection of the X-ray spectra (see below) revealed an essentially 
zero count rate dominated by noise at energies higher than 15 keV.
The mean X-ray count rates averaged over all ten visits to each 
target are $1.30\pm0.32$ ct s$^{-1}$ for DI Lac and 
$0.56\pm0.73$ ct s$^{-1}$ for V841 Oph.  Considering these very 
low count rates, it is not clear whether or not the variability 
seen in the X-ray light curves is real.

\begin{figure}[tb]
\epsscale{1.00}
\plotone{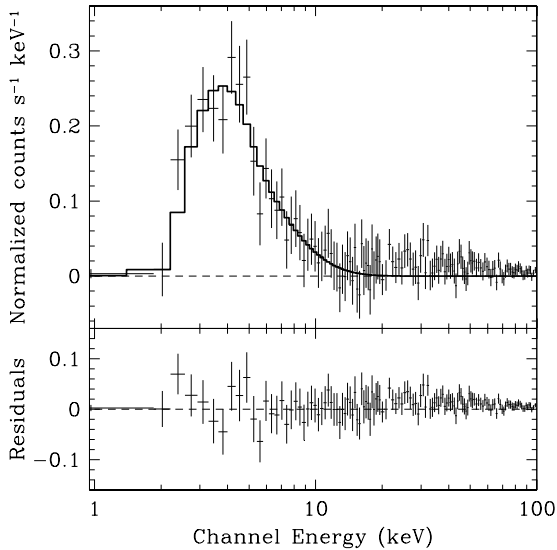}
\epsscale{1.00}
\caption{Top panel shows the combined X-ray spectrum of DI Lac 
(crosses) with the best-fitting bremsstrahlung model modified by 
photoelectric (H column) absorption (solid line).  Bottom panel 
shows the residuals to the model fit.\label{f-DI_modspec}}
\end{figure}

\begin{figure}[tb]
\epsscale{1.00}
\plotone{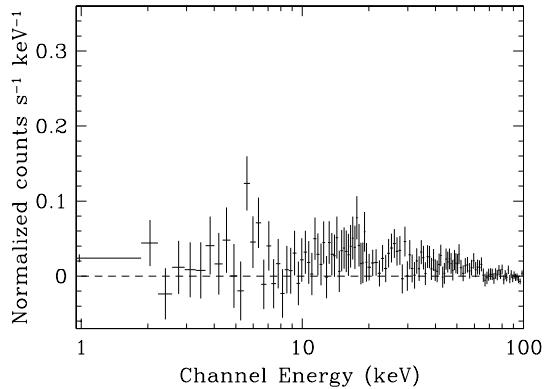}
\epsscale{1.00}
\caption{The combined X-ray spectrum of V841 Oph on the same scale 
as the X-ray spectrum of DI Lac shown in 
Figure \ref{f-DI_modspec}.\label{f-V8_Xspec}}
\end{figure}

Additional analysis and model-fitting (DI Lac only) of the 
X-ray spectra of these novae was performed using the routine 
XSPEC (v11).  Although we extracted the full range of available 
energy channels in the X-ray spectra, only the range 0--15 keV was 
used for fitting the models to the DI Lac data.  The X-ray spectra 
for DI Lac and V841 Oph are shown in Figures \ref{f-DI_modspec} 
and \ref{f-V8_Xspec}.
%%%%%%%%%%%%%%%%%%%%%%%%%%%%%%%%%%%%%%%%%%%%%%%%%%%%%%%%%%%%%%%%%%%
% cp /uw63/hoard/RXTE/DI_Lac/allspec/model_bremss_sm.eps fg8.eps
%%BoundingBox: 20 180 592 765
%%%%%%%%%%%%%%%%%%%%%%%%%%%%%%%%%%%%%%%%%%%%%%%%%%%%%%%%%%%%%%%%%%%
%%%%%%%%%%%%%%%%%%%%%%%%%%%%%%%%%%%%%%%%%%%%%%%%%%%%%%%%%%%%%%%%%%%
% cp /uw63/hoard/RXTE/V841_Oph/allspec/spec_sm.eps fg9.eps
%%BoundingBox: 20 345 592 765
%%%%%%%%%%%%%%%%%%%%%%%%%%%%%%%%%%%%%%%%%%%%%%%%%%%%%%%%%%%%%%%%%%%
The X-ray spectrum of V841 Oph is shown on the same scale as that 
of DI Lac; the former is clearly much weaker than the latter.  We 
attempted to extract a better spectrum of V841 Oph by using only 
the data from the four visits with the largest count rates 
(visits 3, 5, 9, and 11 -- see Table \ref{t-V8_log}).  
Unfortunately, this spectrum was virtually identical to the total 
combined spectrum (i.e.\ flat and featureless), with the exception 
of a small upward offset of the mean count rate, from 
about 0.02 cts s$^{-1}$ keV$^{-1}$ in the total combined spectrum 
to about 0.04 cts s$^{-1}$ keV$^{-1}$ in the ``high'' count rate 
spectrum.  Consequently, no attempt was made to fit a model to the 
spectrum of V841 Oph.  

For DI Lac, we fit three simple models to the X-ray spectrum: 
(1) blackbody, (2) Raymond-Smith thermal plasma, and (3) bremsstrahlung.  
Each model was modified by a multiplicative component representing 
photoelectric (H column) absorption.  More complex models were not 
warranted owing to the small number of counts per energy channel in 
the data.  In all cases, the data were weighted using the 
XSPEC-recommended scheme appropriate for small count numbers, 
$W_{\rm i}=1+(N_{\rm i}+0.75)^{0.5}$ \citep{Gehr86}.  The results 
of these model fits are summarized in the first three rows of 
Table \ref{t-DI_model}.  All of the models produce nearly 
indistinguishable fits to the observed spectrum.
The bremsstrahlung model is formally the best-fitting (with 
lowest reduced $\chi^{2}$ and highest null hypothesis probability) 
and is plotted over the observed spectrum in Figure \ref{f-DI_modspec}.  

All of the models were only able to constrain the hydrogen column 
density to an upper limit; the values of $N_{\rm H}$ quoted in the 
first three rows of Table \ref{t-DI_model} are the upper ends of 
the 1-parameter 90\% confidence intervals.  We used the relation
\begin{displaymath}
N_{\rm H} = E(B-V) \times 5.8\times10^{21} {\rm cm}^{-2} {\rm mag}^{-1}
\end{displaymath}
\citep{Bohl78} to estimate values of $N_{\rm H}$ spanning the 
reddening range given in \S\ref{s-intro}.  We then re-fit the 
models to the X-ray spectrum of DI Lac with only two free 
parameters:\ temperature and normalization.  The results are listed 
in the bottom six rows of Table \ref{t-DI_model}.  While slightly 
different from those obtained from the original model fits, they 
are completely consistent with them.  The bremsstrahlung 
model yields a slightly lower nominal temperature of 
$kT\approx4.1$--4.2 keV (vs.\ $kT=4.5$ keV when $N_{\rm H}$ 
is a free parameter) and is still the best-fitting model for both 
values of $N_{\rm H}$.

We note that the uncertainties quoted in 
Table \ref{t-DI_model} for $kT$ and $C_{\rm norm}$ are 1-parameter 
90\% confidence intervals; the interdependent parameter uncertainties 
will be slightly different.  For example, for the bremsstrahlung 
model (with free $N_{\rm H}$), the XSPEC routine 
steppar\footnote{ This routine iteratively refits the model to the 
data while stepping the values of selected parameters through a 
given range and produces a plot in parameter space of the 
2-dimensional $\chi^{2}$ contours.} gives a 2-parameter 90\% confidence 
interval of 
\begin{displaymath}
kT = 4.5^{+1.9}_{-1.5} {\rm ~keV}
\end{displaymath}
for $kT$ with respect to $N_{\rm H}$, and
\begin{displaymath}
kT = 4.5^{+1.8}_{-1.7} {\rm ~keV}
\end{displaymath}
for $kT$ with respect to $C_{\rm norm}$.  

The X-ray count rates were too low to reliably fit spectrum models 
to any of the ten individual $RXTE$ visits to DI Lac.
We constructed ``low'' and ``high'' state X-ray spectra by combining 
the data from visits with the lowest mean count rates (1, 2, 3, 4, 7) 
and from visits with the highest mean count rates (5, 6, 9) to see 
if there were any differences between the low and high X-ray 
brightness states\footnote{ Visits 8 and 10 were not used.  The 
former was deemed to be an ``intermediate'' brightness state, and 
the latter an anomalously low state.}.  Other than 
an $\approx1$ cts s$^{-1}$ keV$^{-1}$ increase at the peak energy 
of the high state spectrum, the two spectra are essentially 
indistinguishable from each other (and from the combined spectrum 
from all ten visits).  We conclude from this that either the 
spectral energy distribution of the X-ray emitting source in DI Lac 
does not change as the overall brightness of the X-ray source changes, 
or that the variability seen in the X-ray light curve of DI Lac is 
not real (i.e. it is a side-effect of the very low count rates).

\section{Discussion}

\subsection{Optical Photometric Variability}
\label{s-optvar}

Inspection of the RoboScope light curves of DI Lac and V841 Oph 
reveals variability on three distinct time scales.  First, in the 
range 10--100 d, quasiperiods with typical lengths of $\sim30$--50 d 
are present in each of these CVs.  Yet, the presence (or lack) of 
periodic and/or quasiperiodic variability on time scales up to a 
few hundred days in V841 Oph has been contested in the literature.  
(To the best of our knowledge, no extensive database of photometric 
observations of DI Lac has been published prior to this work.)
Using 420 archival visual observations of V841 Oph 
spanning 28 yr, \citet{Dell87} quote an average period of 51.5 d 
with individual cycles ranging from 45 d to 57 d.  \citet{Shar89} 
observed V841 Oph once per night for 31 nights during a 45 night 
interval in 1986 using a CCD camera + $B$ filter, and found 
a period of $\sim40$ d.  All of these values are comparable to 
the $\sim35$--50 d we found for our complete RoboScope light curve of 
V841 Oph, as well as to the range of periods found for individual observing 
seasons in the RoboScope data (see Table \ref{t-periods}).
On the other hand, \citet{Duer92} notes the presence 
of considerable variability in archival light curves of 
V841 Oph \cite[including some of the same data used by][]{Dell87}, 
but finds no evidence for periodicities up to 200 d.

It is quite interesting that the variability time scales in these 
two novae are similar to those that characterize dwarf nova 
outbursts \cite[][ch.\ 3]{Warn95}.  Old novae are presumed to have 
mass transfer rates, $\dot{M}$, above the threshold 
level ($\dot{M}_{\rm crit}$) at which 
the accretion disk thermal instability mechanism operates in dwarf 
novae to produce their outbursts \citep{Osak96}.  However, the 
hibernation theory of cyclic evolution between CV variability 
types \cite[][and references therein]{Sh89}, predicts that 
when $\dot{M}$ eventually decreases back into the instability 
regime some decades after the nova outburst, then dwarf nova 
outbursts should resume.  Both GK Persei \cite[= Nova Per 1901;][]{Bian86} 
and V446 Herculis \cite[= Nova Her 1960;][]{Hone98b} have displayed 
dwarf nova eruptions in their post-nova stages, and similar behavior 
has been suggested for a number of other old novae, 
commencing $\approx$ 50--200 yr after their 
outbursts \cite[][ch.\ 4]{Livi89,Warn95}.
The amplitudes of variability displayed by DI Lac and V841 Oph 
(see Figures \ref{f-DI_lc} and \ref{f-V8_lc}) are smaller than 
those typically observed for dwarf novae ($\Delta V \sim$ 2--4 mag), 
but we cannot rule out the possibility of a disk instability 
mechanism being in operation -- after all, it has been 90 years 
since the nova outburst of DI Lac and over 150 years since that 
of V841 Oph.  Further, the classic thermal instability mechanism 
is not well-explored for high-$\dot{M}$ disks, and other disk 
instabilities might operate as well \cite[e.g.][]{Whit88, Mine93, Godo98}.

The amplitudes of the $\sim30$--50 d oscillations in DI Lac and 
V841 Oph are similar to those of the spaced stunted outbursts seen 
in some novalike CVs \citep{Hone98}.  One of the suggestions for 
stunted outbursts made in \citet{Hone98} is that they are dwarf 
nova eruptions seen against a background of brighter light in the 
system.  If that suggestion is correct, then the oscillations 
reported here \cite[as well as similar 25 d oscillations reported 
for RW Tri,][]{Hone94} may be examples of dwarf nova type outburst 
behavior at relatively short recurrence times, 
$T_{\rm n}\lesssim50$ d \cite[$T_{\rm n}$ is typically $\gtrsim10$ d 
to hundreds of days -- even tens of years in extreme cases -- for 
dwarf novae;][ch.\ 3]{Warn95}.  Among the true dwarf novae, the 
Z Camelopardalis stars have outburst recurrence times of 
$\approx10$--30 d, somewhat shorter than the time scales for the 
oscillations observed in these novae.  The Z Cam stars are thought 
to have high mass transfer rates (near $\dot{M}_{\rm crit}$).  This 
suggests that if the oscillations observed in novae are analogous 
to dwarf nova outbursts, then they may be linked to the presence of 
relatively high mass transfer rates that are, nonetheless, smaller 
than both $\dot{M}_{\rm crit}$ and $\dot{M}$ in the Z Cam stars.

Second, V841 Oph displays a distinct sinusoidal variation with a 
best period of 1800--1900 d (4.9--5.2 yr) or a slightly less 
preferred period of 1250--1300 d (3.4--3.6 yr).  DI Lac does not 
appear to display similar sinusoidal variability, unless it occurs 
with a much longer period than in V841 Oph.  \citet{Bian90} reports 
a long-term period in V841 Oph of 3.4 yr with mean amplitude 
of $\sim0.3$ mag, which was determined from 420 visual observations 
spanning 30 yr \cite[essentially the same early- to mid-20th 
century data utilized by][see below]{Duer92}.  \citet{Rich94} 
re-analyzed the light curves presented by \citet{Bian90} and 
consider the case for a multi-year periodicity in V841 Oph to be 
very weak.  However, with our RoboScope light curve, sinusoidal 
variability on comparable multi-year times scales has now been 
observed in two separate data sets for V841 Oph spanning almost 
80 years of observation.  While this still does not provide firm 
evidence of strictly periodic behavior, it certainly points to the 
presence of a mechanism operating in this CV that modulates its 
brightness on a characteristic time scale of several years.

Photometric variability in old novae on time scales of several years 
has generally been attributed to solar-type magnetic cycles on the 
secondary star that might be able to control the rate of mass loss 
through the inner Lagrangian point \citep{Bian90, Appl92}.  
Although \citet{Rich94} find that no CV studied to date displays 
strictly periodic behavior over multi-year intervals, they also 
note that the observed amplitudes ($\sim0.2$ mag) and ``apparent'' 
time scales of variability of 5--40 yr are ``plausible consequences 
from solar-type magnetic cycles.''  We note that the shorter 
time scale ($\sim$30--50 d) variability discussed above might 
also be linked to modulation of $\dot{M}$ due to secondary star 
magnetic activity.  If solar-type magnetic cycles can 
affect $\dot{M}$ on time scales of years, then starspots induced 
by the magnetic cycle that migrate under the $L_{1}$ point 
might also affect $\dot{M}$ on shorter time scales, in a manner 
suggested by \citet{Livi94} to explain the very low brightness 
states seen in some CVs.

Third, the mean magnitudes of both CVs are changing 
slowly, at rates measured in a few millimagnitudes (mmag) per 
year.  The mean brightness of DI Lac is increasing 
by 4 mmag yr$^{-1}$ since 1990.  \citet{Duer92} analyzed a large 
set of visual, photoelectric, and CCD observations of post-novae 
available in the literature \cite[see references in][]{Duer92}.  
These data cover a large part of the 20th century, from 
the 1920's through the 1980's.  \citet{Duer92} 
summarizes the long-term behavior of DI Lac as exhibiting a steady decline 
of $13\pm1$ mmag yr$^{-1}$ in visual observations obtained 
between 1921 and 1952.  Sparse photoelectric and CCD data 
from 1953 to 1981 suggest a possible small brightness 
increase, which is supported by our recent RoboScope observations.  
For V841 Oph, visual observations from 1919--52 and 
1978--91 yield a brightness decline of $7\pm1$ mmag yr$^{-1}$, 
while ``no definitive brightness decline'' \citep{Duer92} was found in 
photoelectric data from 1954 to 1988.  Our post-1991 observations 
of V841 Oph suggest a trend towards decreasing mean brightness 
comparable to that noted by \citet{Duer92} in the archival 
visual observations, although the obvious sinusoidal variation 
that dominates our light curve at long time scales makes this 
conclusion somewhat suspect.

In the ``classical'' hibernation scenario \citep{Sh89}, novae stay 
bright for $\sim$ 50--300 yr following outburst, gradually 
declining in brightness during this time to a hibernating state 
characterized by low brightness ($M_{V} \sim 10$ or fainter) 
and low mass accretion 
rate ($\dot{M} \lesssim 10^{-12}$ M$_{\odot}$ yr$^{-1}$).  However, 
the photometric record of DI Lac from the 1920's (a decade after 
its outburst) to the 1990's does not show a simple decline in 
brightness.  Instead, the brightness of DI Lac decreased for 
about four decades after outburst, but then apparently leveled 
off for several decades.  In the last decade (covered in our 
RoboScope light curve), the brightness of DI Lac has been 
increasing.  It is possible that we are seeing the influence 
of a mechanism (such as magnetic activity of the secondary 
star -- see above -- and/or accretion-induced irradiation of 
the secondary star) that modulates $\dot{M}$ (and, hence, the 
brightness) in DI Lac over a long-term cycle (with period on 
the order of many decades) that obscures the general decline 
predicted by the hibernation scenario.

\subsection{Optical Spectrum Variability}

The optical spectra of DI Lac and V841 Oph are quite different, 
and this can possibly be ascribed to their difference in post-outburst 
ages.  The spectrum of V841 Oph is similar to those of novalike CVs \cite[e.g.][ch.\ 4]{Warn95} -- this 
suggests that V841 Oph still has a high accretion rate, 
even $\approx150$ yr after outburst.  The narrow emission components 
in DI Lac are suggestive of emission originating from the irradiated 
inner face of the secondary star, while the absorption troughs 
imply the presence of optically thick material in the system.  
The latter feature could be material ejected during the outburst 
of this younger post-nova; the lack of any detected H$\alpha$ 
emission shell \citep{Cohe85} does not preclude the existence 
of denser, non-emitting circumstellar material.  The former 
feature offers a possible explanation for the gradual brightness 
increase seen in our RoboScope light curve of DI Lac (discussed 
in \S\ref{s-optvar}) if the secondary star is being slowly heated 
via irradiation and is, in turn, increasing the rate of mass 
transfer through the $L_{1}$ point.

Although the RoboScope coverage is incomplete, we infer from the 
adjacent data that DI Lac was faint during our July 
spectra and bright during our August spectrum (see 
Figure \ref{f-DI_Xray}).  As mentioned in \S\ref{s-optspec}, the 
only difference between our July and August spectra is that the 
emission cores were somewhat stronger in August (bright state) 
than July (faint state).  This is consistent with the hypothesis 
that these narrow emission features originate on the irradiated 
face of the secondary star if we make the logical assumption 
that the irradiation increases when DI Lac is bright.  It 
does not, however, illuminate the exact mechanism producing the 
irradiation (i.e.\ whether the ``excess'' flux in the high state 
originates near the disk center -- presumably due to a disk 
instability producing increased accretion onto the WD -- or in 
the outer disk -- presumably due to an increase in mass transfer 
through the $L_{1}$ point.)  If we assume that the spectrum 
shown in Figure \ref{f-DI_spec} (from July 27 09:01:47 UT) was 
obtained at an arbitrary orbital phase of 0.0, then the August 17 
spectrum was obtained at a relative orbital phase 
of 0.9 \cite[using $P_{\rm orb} = 0.543773$ d;][]{Ritt98}.  
Thus, these two spectra were obtained at similar orbital phases, 
and we do not expect the difference in the narrow emission 
component strength to be only due to system orientation.

The RoboScope coverage of V841 Oph during our spectroscopic 
(and X-ray) observations is more sparse than for DI Lac, but we 
can infer that the CV was returning to the faint state during our 
May 07 spectrum, was in its faint state during our May 14 and 17 
spectra, and was likely near the bright state during our May 27 
spectrum (see Figure \ref{f-V8_Xray}).  (The two elevated brightness 
points at HJD 2450610 and HJD 2450625 suggest that V841 Oph may 
have returned to a bright state sometime between the low states 
bracketing HJD 2450590 and HJD 2450630.)  If this is the case, 
then the somewhat stronger H$\alpha$ emission of the May 07 and 
May 27 spectra (see Figure \ref{f-V8_spec}) may be linked to the 
bright state of the CV.  As with DI Lac, we calculated relative 
orbital phases for each of these 
spectra \cite[using $P_{\rm orb} = 0.60423$ d;][]{Ritt98}, and 
obtained $\phi = 0.0, 0.6, 0.75, 0.1$ for May 07, 14, 17, 27, 
respectively.  Unfortunately, this casts some doubt on the link 
between H$\alpha$ emission strength and brightness state, since 
both of the bright state spectra (phases 0.0, 0.1) were obtained 
at different orbital phases than the faint state 
spectra (phases 0.6, 0.75).  So, we cannot rule out the influence 
of system orientation effects in these spectra.

\subsection{X-ray Variability}

Unfortunately, because of the weak X-ray emission from these CVs, 
little can be firmly stated about their X-ray variability.  The 
count rates during visits 5 and 6 to DI Lac, which took place when 
we infer from the RoboScope light curve that the CV was in a 
bright state,  are slightly elevated compared to the preceding 
visits.  However, visit 7 also occurred during this optical bright 
state and does not show an elevated count rate.  The mean X-ray 
count rate during each visit to V841 Oph also does not display any 
strong correlation with the corresponding optical state.  
The lack of large changes in X-ray flux or spectrum during the 
optical variations \cite[as usually evident in dwarf 
novae; e.g.][]{Szko99}, argues against a disk instability 
scenario.  Alternatively, this could indicate that
the optical brightness is determined by activity in the 
outer disk only and, therefore, is not reflected in the X-ray 
behavior; however, the X-ray count rates are too low (and their 
corresponding error bars too large) to make any firm conclusions.

\section{Conclusions}

Our long-term optical light curves of the novae DI Lac and V841 Oph 
obtained with RoboScope reveal quasiperiodic variability with a 
characteristic time scale of $\sim30$--50 d in both CVs.  In addition, 
the light curve of V841 Oph displays evidence for sinusoidal 
variability with a period of 3.5--5 yr.  The latter cannot be said 
to be strictly periodic since our data set covers only $\approx1.5$ 
cycles; however, when this detection is added to past reports of 
multi-year periodicities in V841 Oph that have been reported in the 
literature, a strong case can be made for the presence of repeating 
multi-year variability with a preferred time scale in this CV.  The 
most likely origin of such behavior is a solar-type magnetic cycle 
of the secondary star that modulates the rate of mass transfer 
through the $L_{1}$ point.  If this is the case, then the 
shorter $\sim30$--50 d quasiperiodic variability in this system 
might also be related to the magnetic activity on the secondary 
star; for example, due to starspots that migrate under the $L_{1}$ 
point and temporarily throttle mass transfer.  DI Lac does not 
show evidence for cyclic multi-year variability, unless it occurs 
with a much longer period than the length of our RoboScope 
coverage ($\gtrsim10$ yr).  This casts some doubt on the origin 
of the $\sim30$--50 d quasiperiodic variability as a facet of 
the secondary star's magnetic activity, since both V841 Oph and 
DI Lac show the $\sim30$--50 d quasiperiodic behavior, but only 
V841 Oph displays evidence for a multi-year solar-type magnetic 
cycle on the secondary star.  On the other hand, the $\sim30$--50 d 
time scale is very reminiscent of that expected for dwarf nova type 
behavior (although the amplitude of variability in these novae is 
smaller than in typical dwarf nova outbursts).  This raises the 
possibility that this variability is caused by a disk instability 
(either the thermal disk instability that leads to dwarf nova 
outbursts operating at a low level or some other form of disk 
instability).  

The X-ray spectrum of DI Lac is fit almost equally well by the 
three models we tried: a simple blackbody, a Raymond-Smith thermal 
plasma, and bremsstrahlung emission.  More complicated models 
(e.g. involving multiple components, lines, etc.) are unwarranted 
due to the low X-ray flux.  In addition to being the most 
physically plausible X-ray emission emission mechanism in a 
(non-magnetic) CV, the bremsstrahlung model is formally the 
best-fitting.  Our X-ray spectrum model fit 
parameters ($kT\sim4$ keV) are consistent with those obtained 
from $ROSAT$ X-ray spectra of a sample of 37 disk-accreting 
CVs \citep{Rich96}.  Unfortunately, the X-ray count rates in 
both of these systems were too low for any conclusions to be 
made about their time-resolved X-ray behavior.  X-ray 
observations sample the innermost disk region, and we would 
expect to see different time-resolved behavior as the novae 
go into their optically bright state if the transition is 
triggered by a disk instability vs.\ a change in $\dot{M}$ 
from the secondary star.  Additional time-resolved X-ray 
observations using the more sensitive {\em Chandra} 
and/or {\em XMM} X-ray satellites may be necessary to 
illuminate the inner workings of these old novae.

\acknowledgements

PS, DWH, and VD acknowledge support from NASA grant NAG5-4791.  
DWH thanks the NOAO librarian Mary Guerrieri for her valuable 
assistance locating several papers cited herein.
This research made use of NASA's Astrophysics Data System 
Abstract Service and the SIMBAD database operated by CDS, 
Strasbourg, France.

\singlespace

%%%%%%%%%%%%%%%%%%%%%%%%%%%%%%%%%%%%%%%%%%%%%%%%%%%%%%%%%%%%%%%%%%%%%%%%%%%%%
\begin{deluxetable}{llcccl}
\tablewidth{0pt}
\tablecaption{Log of Observations for DI Lac\label{t-DI_log}}
\tablehead{
\colhead{UT Date and Time\tablenotemark{a}} & 
\colhead{HJD\tablenotemark{a}} & 
\colhead{Observatory} & 
\colhead{Mode\tablenotemark{b}} &
\colhead{Exposure} &
\colhead{Comment} \\
\colhead{ } & 
\colhead{ } & 
\colhead{ } & 
\colhead{ } &
\colhead{or GTI (s)} &
\colhead{ }  
}
\startdata
1990 Nov 12 & 2448207.7 & RoboScope & P & \phn240\tablenotemark{c} & starting date \\
1997 Jul 25 10:43:44 & 2450654.94822 & RXTE & X & 2528 & ID 20037-02-01-00 \\
1997 Jul 27 09:01:47 & 2450654.87744 & APO & S & \phn900 & \nodata \\
1997 Jul 27 09:16:53 & 2450654.88792 & APO & S & \phn737 & \nodata \\
1997 Jul 30 04:16:00 & 2450659.67921 & RXTE & X &  \phn944 & ID 20037-02-02-00 \\
1997 Aug 03 04:11:12 & 2450663.68424 & RXTE & X & 1536 & ID 20037-02-03-00 \\
1997 Aug 07 04:12:00 & 2450667.67683 & RXTE & X & 1680 & ID 20037-02-04-00 \\ 
1997 Aug 12 03:42:40 & 2450672.65668 & RXTE & X &  \phn592 & ID 20037-02-05-00 \\
1997 Aug 17 03:42:18 & 2450677.65666 & APO & S & \phn900 & \nodata \\
1997 Aug 17 03:45:36 & 2450677.65893 & RXTE & X & 1024 & ID 20037-02-06-00 \\
1997 Aug 23 03:46:40 & 2450683.65991 & RXTE & X &  \phn832 & ID 20037-02-07-00 \\
1997 Aug 27 03:46:56 & 2450687.66024 & RXTE & X &  \phn736 & ID 20037-02-08-00 \\
1997 Sep 01 08:12:48 & 2450692.84503 & RXTE & X & 1616 & ID 20037-02-09-00 \\
1997 Sep 03 05:56:28 & 2450694.74974 & WIYN & S & \phn900 & No flux calibration \\
1997 Sep 04 04:33:17 & 2450695.69269 & WIYN & S & \phn900 & No flux calibration \\
1997 Sep 06 08:08:48 & 2450697.84239 & RXTE & X & 1776 & ID 20037-02-10-00 \\
1998 Nov 30 & 2451147.6 & RoboScope & P & \phn240\tablenotemark{c} & ending date
\enddata
\tablenotetext{a}{ At mid-point of exposure or Good-Time-Interval (GTI)}
\tablenotetext{b}{ P = ground-based photometry; S = ground-based spectroscopy; 
X = X-ray}
\tablenotetext{c}{ 1--2 measurements night$^{-1}$}
\end{deluxetable}
%%%%%%%%%%%%%%%%%%%%%%%%%%%%%%%%%%%%%%%%%%%%%%%%%%%%%%%%%%%%%%%%%%%%%%%%%%%%%

%%%%%%%%%%%%%%%%%%%%%%%%%%%%%%%%%%%%%%%%%%%%%%%%%%%%%%%%%%%%%%%%%%%%%%%%%%%%%
\begin{deluxetable}{llcccl}
\tablewidth{0pt}
\tablecaption{Log of Observations for V841 Oph\label{t-V8_log}}
\tablehead{
\colhead{UT Date and Time\tablenotemark{a}} & 
\colhead{HJD\tablenotemark{a}} & 
\colhead{Observatory} & 
\colhead{Mode\tablenotemark{b}} &
\colhead{Exposure} &
\colhead{Comment} \\
\colhead{ } & 
\colhead{ } & 
\colhead{ } & 
\colhead{ } &
\colhead{or GTI (s)} &
\colhead{ }  
}
\startdata
1991 May 31 & 2448407.7 & RoboScope & P & \phn240\tablenotemark{c} & starting date \\
1997 Apr 30 16:10:57 & 2450569.17897 & RXTE & X & 1936 & ID 20037-01-01-00 \\ 
1997 May 05 12:50:25 & 2450574.03997 & RXTE & X & 1856 & ID 20037-01-02-00 \\
1997 May 07 06:30:32 & 2450575.77120 &  APO & S & \phn600 & \nodata \\
1997 May 10 20:59:13 & 2450579.37966 & RXTE & X & 1936 & ID 20037-01-03-00 \\
1997 May 14 06:37:42 & 2450582.77618 &  APO & S & \phn600 & \nodata \\
1997 May 15 21:12:33 & 2450584.38911 & RXTE & X & 1408 & ID 20037-01-04-00 \\
1997 May 17 09:35    & 2450585.899   & WIYN & S & 4500 & 3 combined spectra \\
1997 May 20 07:58:41 & 2450588.83795 & RXTE & X & 1792 & ID 20037-01-05-00 \\
1997 May 26 06:28:11 & 2450594.77523 & RXTE & X & 1872 & ID 20037-01-06-00 \\
1997 May 27 06:22:21 & 2450595.76552 &  APO & S & \phn600 & \nodata \\
1997 May 30 05:06:57 & 2450598.71888 & RXTE & X & 1648 & ID 20037-01-07-00 \\
1997 Jun 04 22:57:37 & 2450604.46244 & RXTE & X & \phn832 & ID 20037-01-08-00 \\
1997 Jun 09 06:43:45 & 2450608.78613 & RXTE & X & 2160 & ID 20037-01-09-00 \\
1997 Jun 18 16:24:33 & 2450618.18935 & RXTE & X & 1376 & ID 20037-01-11-00\tablenotemark{d} \\
1998 Sep 04 & 2451060.5 & RoboScope & P & \phn240\tablenotemark{c} & ending date
\enddata
\tablenotetext{a}{ At mid-point of exposure or Good-Time-Interval (GTI)}
\tablenotetext{b}{ P = ground-based photometry; S = ground-based spectroscopy; 
X = X-ray}
\tablenotetext{c}{ 1--2 measurements night$^{-1}$}
\tablenotetext{d}{ Our tenth RXTE observation of V841 Oph on 
1997 Jun 14 UT (ID 20037-01-10-00) was unusable due to a software event 
that shut down all of the detectors during the slew to the target.}
\end{deluxetable}
%%%%%%%%%%%%%%%%%%%%%%%%%%%%%%%%%%%%%%%%%%%%%%%%%%%%%%%%%%%%%%%%%%%%%%%%%%%%%

%%%%%%%%%%%%%%%%%%%%%%%%%%%%%%%%%%%%%%%%%%%%%%%%%%%%%%%%%%%%%%%%%%%%%%%%%%%%%
\begin{deluxetable}{ccc}
\tablewidth{0pt}
\tablecaption{Periods in Optical Light Curves\label{t-periods}}
\tablehead{
\colhead{Observing} &
\colhead{DI Lac} &
\colhead{V841 Oph} \\
\colhead{Season} &
\colhead{Period (d)} &
\colhead{Period (d)}
}
\startdata
1 & 40 & 35 \\
2 & 29 & 36 \\
3 & 39 & 69 \\
4 & 31 & 40 \\
5 & 38 & 42 \\
6 & 75,40 & 49 \\
7 & 80,28 & 54 \\
8 & 31,21 & 48 \\
combined & 43,37 & 49,36
\enddata
\end{deluxetable}
%%%%%%%%%%%%%%%%%%%%%%%%%%%%%%%%%%%%%%%%%%%%%%%%%%%%%%%%%%%%%%%%%%%%%%%%%%%%%

%%%%%%%%%%%%%%%%%%%%%%%%%%%%%%%%%%%%%%%%%%%%%%%%%%%%%%%%%%%%%%%%%%%%%%%%%%%%%
\begin{deluxetable}{lcccccc}
\tablecolumns{7}
\tablewidth{0pt}
\tablecaption{X-ray Spectrum Models for DI Lac\label{t-DI_model}}
\tablehead{
\colhead{Model} & 
\colhead{Reduced$\chi^{2}$} & 
\colhead{$P_{\rm 0}$\tablenotemark{b}} &
\colhead{H Column} &
\colhead{2--15 keV} &
\multicolumn{2}{c}{Parameters\tablenotemark{c}} \\
\colhead{ } & 
\colhead{(47 dof)\tablenotemark{a}} & 
\colhead{ } &
\colhead{Density, $N_{\rm H}$} &
\colhead{Model Flux} &
\colhead{$kT$} &
\colhead{$C_{\rm norm}$} \\
\colhead{ } & 
\colhead{ } & 
\colhead{ } &
\colhead{($10^{22}$ cm$^{-2}$)} &
\colhead{(erg s$^{-1}$ cm$^{-2}$)} &
\colhead{(keV)} &
\colhead{ }
}
\startdata
%%%%%%%%%%%%%%%%%%%%%%%%%%%%%%%%%%%%%%%%%%%%%%%%%%%%%%%%%%%%%%%%%%%%%%%%%
%%%%% 3 decimal places
%Blackbody & 0.705 & 0.937  & $\le1.23$ & $2.18\times10^{-12}$ & 
%$1.12^{+0.10}_{-0.16}$ & $3.03^{+0.59}_{-0.18}\times10^{-5}$ \\[4pt]
%
%Raymond-Smith\tablenotemark{d} & 0.617 & 0.982 & $\le3.34$ & 
%$2.64\times10^{-12}$ & $3.16^{+0.73}_{-1.06}$ & 
%$3.38^{+3.49}_{-0.58}\times10^{-3}$ \\[4pt]
%
%Bremsstrahlung & 0.484 & 0.999 & $\le1.83$ & $2.74\times10^{-12}$ & 
%$4.47^{+1.26}_{-1.33}$ & $1.10^{+0.68}_{-1.10}\times10^{-3}$ \\[2pt]
%
%\sidehead{H Column Density fixed from $E(B-V)=0.15$:}
%
%Blackbody & 0.683 & 0.954 & 0.087 & $2.21\times10^{-12}$ & 
%$1.08^{+0.14}_{-0.13}$ & $3.15^{+0.42}_{-0.29}\times10^{-5}$ \\[4pt]
%
%Raymond-Smith\tablenotemark{d} & 0.604 & 0.986 & 0.087 & $2.67\times10^{-12}$ 
%& $2.91^{+0.95}_{-0.51}$ & $3.74^{+0.86}_{-0.91}\times10^{-3}$ \\[4pt]
%
%Bremsstrahlung & 0.471 & 0.999 & 0.087 & $2.75\times10^{-12}$ & 
%$4.19^{+1.48}_{-1.01}$ & $1.19^{+0.42}_{-0.27}\times10^{-3}$ \\[2pt]
%
%\sidehead{H Column Density fixed from $E(B-V)=0.41$:}
%
%Blackbody & 0.689 & 0.951  & 0.238 & $2.28\times10^{-12}$ & 
%$1.07^{+0.14}_{-0.12}$ & $3.30^{+0.31}_{-0.40}\times10^{-5}$ \\[4pt]
%
%Raymond-Smith\tablenotemark{d} & 0.606 & 0.986 & 0.238 & $2.66\times10^{-12}$ 
%& $2.86^{+0.90}_{-0.49}$ & $3.85^{+0.87}_{-0.93}\times10^{-3}$ \\[4pt]
%
%Bremsstrahlung & 0.475 & 0.999 & 0.238 & $2.73\times10^{-12}$ & 
%$4.12^{+1.45}_{-0.99}$ & $1.23^{+0.44}_{-0.29}\times10^{-3}$ 
%%%%%%%%%%%%%%%%%%%%%%%%%%%%%%%%%%%%%%%%%%%%%%%%%%%%%%%%%%%%%%%%%%%%%%%%%%
Blackbody & 0.705 & 0.937  & $\le1.2$ & $2.2\times10^{-12}$ & $1.1^{+0.1}_{-0.2}$ & $3.0^{+0.6}_{-0.2}\times10^{-5}$ \\[4pt]

Raymond-Smith\tablenotemark{d} & 0.617 & 0.982 & $\le3.3$ & $2.6\times10^{-12}$ & $3.2^{+0.7}_{-1.1}$ & $3.4^{+3.5}_{-0.6}\times10^{-3}$ \\[4pt]

Bremsstrahlung & 0.484 & 0.999 & $\le1.8$ & $2.7\times10^{-12}$ & $4.5^{+1.3}_{-1.3}$ & $1.1^{+0.7}_{-1.1}\times10^{-3}$ \\[2pt]

\sidehead{H Column Density fixed from $E(B-V)=0.15$:}

Blackbody & 0.683 & 0.954 & 0.09 & $2.2\times10^{-12}$ & $1.1^{+0.1}_{-0.1}$ & $3.1^{+0.4}_{-0.3}\times10^{-5}$ \\[4pt]

Raymond-Smith\tablenotemark{d} & 0.604 & 0.986 & 0.09 & $2.7\times10^{-12}$ & $2.9^{+0.9}_{-0.5}$ & $3.7^{+0.9}_{-0.9}\times10^{-3}$ \\[4pt]

Bremsstrahlung & 0.471 & 0.999 & 0.09 & $2.7\times10^{-12}$ & $4.2^{+1.5}_{-1.0}$ & $1.2^{+0.4}_{-0.3}\times10^{-3}$ \\[2pt]

\sidehead{H Column Density fixed from $E(B-V)=0.41$:}

Blackbody & 0.689 & 0.951  & 0.24 & $2.3\times10^{-12}$ & $1.1^{+0.1}_{-0.1}$ & $3.3^{+0.3}_{-0.4}\times10^{-5}$ \\[4pt]

Raymond-Smith\tablenotemark{d} & 0.606 & 0.986 & 0.24 & $2.7\times10^{-12}$ & $2.9^{+0.9}_{-0.5}$ & $3.8^{+0.9}_{-0.9}\times10^{-3}$ \\[4pt]

Bremsstrahlung & 0.475 & 0.999 & 0.24 & $2.7\times10^{-12}$ & $4.1^{+1.4}_{-1.0}$ & $1.2^{+0.4}_{-0.3}\times10^{-3}$ 

\enddata
\tablenotetext{a}{ 48 dof for fixed $N_{\rm H}$ models.}
\tablenotetext{b}{ Null Hypothesis Probability: the probability of 
obtaining a value of $\chi^{2}$ greater than or equal to the 
observed $\chi^{2}$ if the model is correct.  A value close to 1.0 
indicates a good fit.}
\tablenotetext{c}{ Quoted uncertainties are 1-parameter 90\% confidence 
intervals.  $T$ is the plasma temperature for the bremsstrahlung and 
Raymond-Smith plasma models.  In all models, $C_{\rm norm}$ is a 
normalization constant.}
\tablenotetext{d}{ Additional model parameters of redshift and abundances 
were fixed at 0.0 and the values from \citet{Feld92}, respectively.}
\end{deluxetable}
%%%%%%%%%%%%%%%%%%%%%%%%%%%%%%%%%%%%%%%%%%%%%%%%%%%%%%%%%%%%%%%%%%%%%%%%%%%%%

\end{document}